\documentclass[reprint,aps,pra,superscriptaddress]{revtex4-2}
\usepackage[utf8]{inputenc}

\usepackage{xcolor}

\definecolor{algogreen}{RGB}{46, 106, 107}

\usepackage[colorlinks=true,%
bookmarks=false,%
linkcolor=algogreen,%
urlcolor=algogreen,%
citecolor=algogreen,%
breaklinks]{hyperref}

\usepackage{graphicx}

\begin{document}

\title{Quantum network medicine: rethinking medicine\\with network science and quantum algorithms}
\author{Sabrina Maniscalco}
\affiliation{Algorithmiq Ltd, Kanavakatu 3C, FI-00160 Helsinki, Finland}
\email{info@algorithmiq.fi}
\homepage{www.algorithmiq.fi}
\affiliation{QTF Centre of Excellence, Department of Physics, Faculty of Science, University of Helsinki, Finland}
\affiliation{InstituteQ - the Finnish Quantum Institute, University of Helsinki, Finland}
\affiliation{QTF Centre of Excellence, Department of Applied Physics, Aalto University, FI-00076 Aalto, Finland} 
\author{Elsi-Mari Borrelli}
\author{Daniel Cavalcanti}
\author{Caterina Foti}
\author{Adam Glos}
\author{Mark Goldsmith}
\author{Stefan Knecht}
\author{Keijo Korhonen}
\author{Joonas Malmi}
\author{Anton Nykänen}
\author{Matteo A. C. Rossi}
\author{Harto Saarinen}
\author{Boris Sokolov}
\author{N. Walter Talarico}
\author{Jussi Westergren}
\author{Zoltán Zimborás}
\affiliation{Algorithmiq Ltd, Kanavakatu 3C, FI-00160 Helsinki, Finland}
\email{info@algorithmiq.fi}
\homepage{www.algorithmiq.fi}
\author{Guillermo García-Pérez}
\affiliation{Algorithmiq Ltd, Kanavakatu 3C, FI-00160 Helsinki, Finland}
\email{info@algorithmiq.fi}
\homepage{www.algorithmiq.fi}

\date{\today}

\begin{abstract}
Scientific and technological advances in medicine and systems biology have unequivocally shown that health and disease must be viewed in the context of the interplay among multiple molecular and environmental factors.
Understanding the effects of molecular and cellular interconnections on disease progression may lead to the identification of novel disease genes and pathways, and, hence, influence precision diagnostics and therapeutics.
To accomplish this goal, the emerging field of network medicine applies network science approaches to investigate disease pathogenesis, integrating information from relevant Omics databases, including protein–protein interaction, correlation-based, gene regulatory, and Bayesian networks.
However, this approach requires analysing and computing large amounts of data.
Moreover, if we are to search efficiently for new drugs and new drug combinations, there is a pressing need for computational methods that could allow us to access the immense chemical compound space that remains largely unexplored.
Finally, at the microscopic level, drug-target chemistry simulation is ultimately a quantum problem, and, hence, requires a quantum solution.
As we will discuss, quantum computing may be a key ingredient in enabling the full potential of network medicine.
We propose to combine network medicine and quantum algorithms in a novel research field, quantum network medicine, to lay the foundations of a new era of disease mechanism, prevention, and treatment.
\end{abstract}

\maketitle

\section{Background}
Remarkable progress in science and technology has profoundly changed our society in the last decades.
The improvement in measurement techniques and instrumentation has led to tremendous advances in natural sciences, from physics to chemistry, to the life sciences.
These advances cover phenomena at all scales and of varying complexity, from the detection of gravitational waves to the control of microscopic constituents of matter, from high-throughput DNA sequencing to quantitative protein analysis.
As a consequence, our ability to understand, model, and control complex phenomena has led to new, emerging disciplines and technologies.
In this white paper, we focus on two such examples, quantum computing~\cite{noauthor_40_2022} and network medicine~\cite{loscalzo_human_2007}, that combined may lead to foundational breakthroughs and disruptive impact both in physics and medicine.

\begin{figure*}[t]
    \includegraphics[width=0.9 \textwidth]{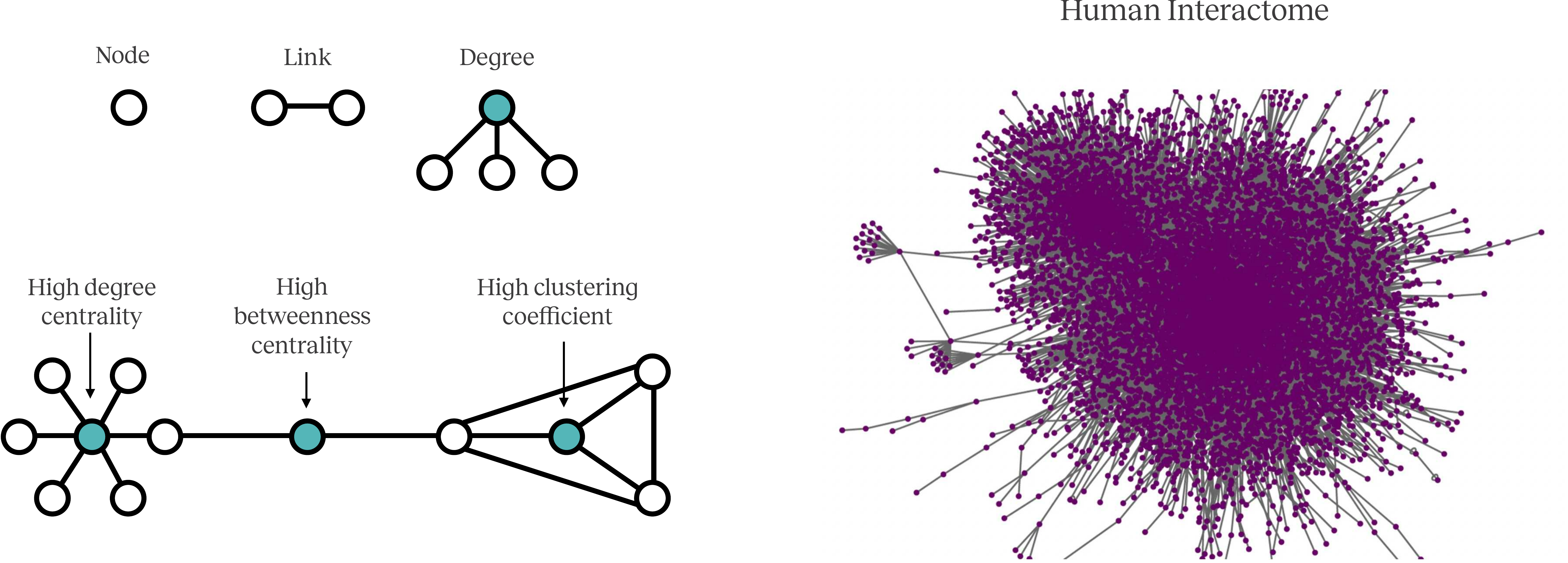}
    \caption{A network is described by a graph $G = (N, E)$ with $N$ the set of nodes, and $E$ the set of edges or links;
    the degree of a node is defined as the number of links incident to the node (equal to 3 for the red node in the top row of the left figure).
    A hub is a node connected to a large fraction of the system, as the node with high degree centrality in the figure.
    A node has high betweenness centrality if many of the shortest paths between pairs of nodes in the network pass through it.
    Clustering coefficient is a measure of the degree to which nodes in a graph tend to cluster together forming connected triples of nodes.
    The network in the right figure is the Human Reference Interactome (HuRI) map, the largest of its kind, charting 52,569 interactions between 8,275 human proteins~\cite{luck_reference_2020}.}
    \label{fig1}
 \end{figure*}

This paper summarises the main achievements and challenges of each of these fields and discusses how quantum computers may provide the needed computational boost to network medicine, while network medicine may provide several illustrative use cases of practical relevance to quantum computing.
We call the combination and synergy between these two fields quantum network medicine.
It is worth emphasizing from the start that we by no means imply that the only boost provided by quantum computers is in terms of speed of calculations of the typical computations performed in network medicine and systems biology.
In fact, we will argue that quantum network medicine is a multi-scale framework encompassing data-driven approaches exploiting Omic databases (large scale), protein folding and modelling (meso-scale), and quantum simulations of chemical reactions, such as protein-ligand binding (microscopic scale).

Changes of paradigm often come from interdisciplinary endeavours, the consequent knowledge exchange, and cross-fertilisation, including the realisation that tools and approaches from one discipline can be profoundly useful in another.
Fresh perspectives on a problem using a different mindset may unlock roadblocks that have existed for decades.
We champion the creation of a new multidisciplinary community and ecosystem, bringing together experts on quantum science and technology, network science, chemistry, and life sciences to tackle some of the grand challenges of our society and create long-lasting impact.

\section{Network Medicine}
Network medicine stems from the realisation that scientific reductionism is inadequate for understanding the mechanisms underlying complex diseases.
Consequently, a more holistic approach is needed to increase the efficacy of prevention strategies, to tailor precise therapies, and to modernise drug discovery and development (DDD). \cite{Schmidt2022-vr}.
The term was coined in 2007 by Albert-László Barabási, Isaac Kohane, and Joseph Loscalzo, who proposed using network science approaches in systems biology~\cite{loscalzo_human_2007}.
In the following, we provide a short summary of the motivation, methodology, successes, and challenges of this emerging field.

\subsection{Motivation}
In the great majority of cases, the origin of a disease cannot be simply traced back to an abnormality of a single gene in a human cell.
On the contrary, interdependencies between different molecular components are reflected in changes in both intracellular and intercellular networks linking tissues and organ systems.
In order to gain understanding of these complex mechanisms, network medicine represents Omics Big Data, which includes genetics, epigenetics, transcriptomics, metabolomics, and proteomics, in terms of networks composed of nodes and links.
Examples of molecular networks are protein-protein interaction (PPI) networks, the nodes of which are proteins that are linked to each other by binding interactions, and metabolic networks, where the nodes are metabolites and the links represent shared biochemical (enzymatic) reactions.
Some important concepts generally used in network science are depicted in Fig.~\ref{fig1}, together with a visual representation of one of the most important molecular networks, the human PPI network. 

Recent advances in network theory have demonstrated that networks operating in biological, technological or social systems (real networks) are characterised by a similar set of underlying principles and properties that make them very different from regular structures and, yet, not completely random.
For example, real networks are often scale-free, that is, they are characterised by very broad degree distributions (the degree of a node is defined as the number of neighbours it has in the network, see Fig.~\ref{fig1}) so that most nodes have very small degree while a few nodes, called hubs, are connected to a large fraction of the nodes in the system~\cite{barabasi_emergence_1999}.
Generally, this makes real networks resilient to the the loss of many nodes, except for the removal of hubs, which can cause them to collapse.
Real networks are also typically small-world, meaning that most pairs of nodes are very few edges away in the network~\cite{watts_collective_1998}, and have very high levels of clustering, that is, there are many more connected triples of nodes than one would expect from a purely random distribution of links~\cite{serrano_self-similarity_2008}.
Complex network theory thus studies complex systems from the point of view of these network representations.
In particular, its main goals are to provide a holistic description of a given system in terms of statistical properties of its graph representation, to model and explain the origin of the uncovered structural properties and, ultimately, to understand and predict the impact of such network structure on the global system's behaviour.

\begin{figure*}[t]
    \includegraphics[width=0.9 \textwidth]{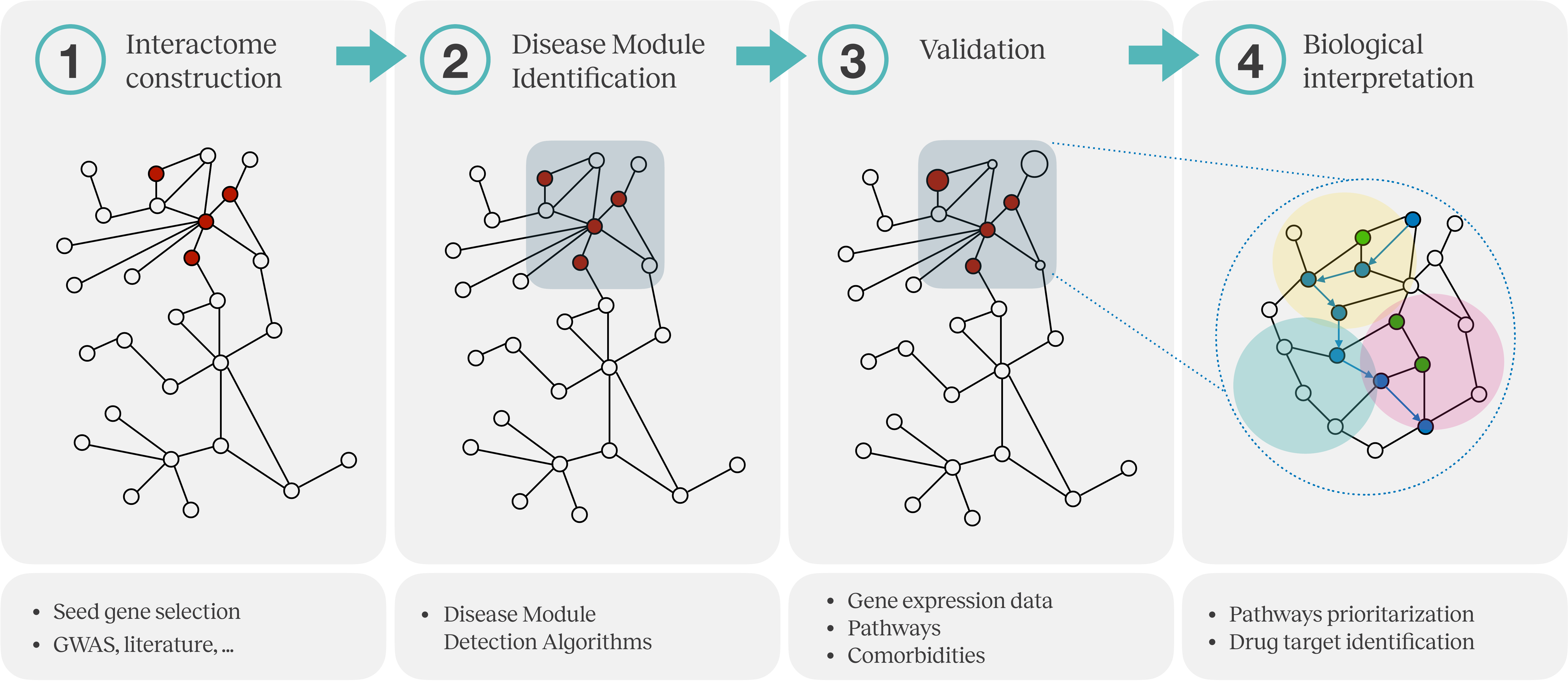}
    \caption{Main steps in the identification, validation, and use of disease modules.
    1) Upon construction of the interactome, GWAS or experimental data are used to identify disease genes to be used as seeds (marked in red).
    2) The disease module is then determined by analysing the neighbourhood of the seed nodes in the interactome by, for example, using diffusion-based methods.
    3) The resulting putative module is later validated using additional data, such as gene co-expression data or disease comorbidities.
    4) The validated disease module is then analysed to understand the disease mechanisms and, ultimately, to identify potential drug targets.}
    \label{fig:fig2}
 \end{figure*}

In the network medicine scenario, network properties are used to identify topological indicators for the location of disease genes within the network of molecular interactions---the interactome.
For example, it was hypothesised that hub proteins may have a special biological role, given that one would expect the removal of a hub to affect the system more than the absence of non-hub proteins and, indeed, early results suggested a direct correlation between node degree and essentiality (i.e., how critical a node is for survival)~\cite{jeong_lethality_2001}.
While subsequent results challenge this relationship~\cite{coulomb_gene_2005}, the hypothesis that protein function is closely related to network topology has been strengthened by studies using centrality measures, that is, quantities aimed at gauging nodes' relevance beyond their degree.
It has been found that nodes with high betweenness centrality, meaning that many of the shortest paths between pairs of nodes in the network pass through them, tend to be more essential than hubs~\cite{hwang_identification_2008, yu_importance_2007}.
The elimination of non-hub proteins with high bridging centrality (that is, nodes connecting modular communities within the network), instead, has been found less likely to be lethal~\cite{hwang_identification_2008}.
Given their key role in connecting network modules and their low lethality, bridging nodes have been suggested as potential drug targets.

Other compelling theoretical and experimental studies support the hypothesis that network topology is an essential feature in the emergent system function of the protein when it is perturbed.
For this reason, network theory provides a useful basis for developing methods to prioritise drug targets and combinations of targets~\cite{hopkins_network_2008}.

\subsection{Methodology}
Network medicine integrates approaches to Omics Big Data and combines them with computational biology tools.
In the following, we briefly survey some of the analytical methods and concepts used in this emerging discipline.

\subsubsection{Identifying disease modules}\label{sec:DiseaseModule}
One of the assumptions of network medicine, the `local' hypothesis, asserts that if a gene product or protein is involved in a specific biochemical process or disease mechanism, its direct interactors, which are found in its network-based vicinity, might also have some role in the same biochemical process.
Similarly, proteins that are involved in the same disease show a high propensity to interact with each other.
Based on these observations, we expect that the proteins governing each disease can be localized in a certain neighbourhood of the interactome.
This neighbourhood is called a \textit{disease module}~\cite{barabasi_network_2011}, where in network science module (or community) indicates a dense subgraph on the network that often represents a set of nodes having a joint role or contributing to some common function.
Node or edge perturbations of a disease module (such as mutations, deletions, or expression changes) are linked to a particular disease phenotype.

It is important to distinguish disease modules from two other types of modules or communities, \textit{topological modules} and \textit{functional modules}.
The former are locally dense neighbourhoods in a network and can be identified using community detection algorithms~\cite{fortunato_community_2016} that are blind to their function or role in specific diseases of individual nodes.
As for the latter, functional modules are inherited from systems biology, where a module represents a group of molecules that interact with each other to achieve some common functional response.

It is believed that these three types of modules are interrelated.
More specifically, it is often assumed that cellular components of a topological module have closely related functions.
Thus, they will likely partly overlap with (or be part of) a functional module.
In turn, since diseases are caused by the breakdown of functional modules, topological modules may be connected to disease modules as well.
However, one should remember that, while each disease has its own module, a gene, protein or metabolite can be implicated in several disease modules.
Hence, modules related to different diseases are not necessarily separated, but can, instead, be connected if they share some mechanistic pathways.

Determining the disease module for a given pathophenotype of interest is crucial in order to direct experimental work towards understanding of the disease mechanism, and can, therefore, also play an important role in drug development.
The identification of the module typically relies on the assumption that the disease-related proteins are localised in the interactome (that is, the average distance between pairs of them in the network is shorter than between randomly chosen pairs in the whole network).
Guided by this principle, many approaches have been proposed for identifying and characterizing disease modules~\cite{lee_network_2019}.

Once the interactome is available, the starting point is the identification of so-called seed disease-genes based on, e.g., experimental evidence or genome-wide association studies (GWAS)~\cite{ghiassian_disease_2015,jia_dmgwas_2011,petti_connectivity_2020}.
The disease module is then constructed by selecting relevant genes in the vicinity of the seed disease-genes according to different criteria, ranging from a simple nearest-neighbours choice to more elaborate strategies based on random walk dynamics~\cite{erten_dada_2011,navlakha_power_2010}.

The latter methods, which tend to give the best performance, are often referred to as \textit{diffusion-based}, and they aim at taking into account pathway-related properties of the system.
Essentially, these techniques base their predictions on the simulation of a walker diffusing along the links of the interactome, randomly hopping between neighbouring nodes (or back to the initial seed node), with the prediction of whether other proteins belong to the disease module depending on their probability of being visited by the walker.

The resulting predictions are finally validated, for instance, by showing that the genes within the identified module have correlated expression patterns or related functions.
Finally, the validated disease module can be used to extract biological insights on the disease mechanism and, ultimately, to identify potential drug targets (see Fig.~\ref{fig:fig2}).

\subsubsection{Disease networks}

As previously discussed, the fact that proteins can belong to multiple disease modules implies that diseases cannot be considered independent of one another.
This interdependence between diseases can also be represented in terms of complex networks, the so-called \textit{disease networks}~\cite{goh_human_2007}, in which nodes correspond to diseases and links to molecular relationships between the disease-associated cellular components.
Constructing accurate disease networks by identifying such connections between diseases could result in novel strategies for disease prevention, diagnosis, and treatment.
For example, an immediate application is drug repurposing, i.e., using approved drugs to treat other diseases than those for which they were approved.

\subsection{Successes and challenges in network medicine}
A number of successful applications of network medicine analyses have been reported to date, including the identification of a putative drug target in pulmonary arterial hypertension~\cite{samokhin_nedd9_2018} and in coronary heart disease~\cite{wang_network-based_2018}.
In the case of coronary heart disease, an approach using the distance between disease proteins and drug targets in the human PPI revealed the potential of network medicine for drug repurposing~\cite{cheng_network-based_2018}.
Other examples include the identification of potentially crucial nodes in the pathogenesis of diabetes mellitus~\cite{sharma_disease_2015}, and even of potential genes and pathways shared between phenotypically distinct diseases such as chronic obstructive pulmonary disease and idiopathic pulmonary fibrosis~\cite{halu_exploring_2019}.

Despite this substantial progress in a relatively short period since network medicine's first inception, some important knowledge gaps and challenges have limited its impact on clinical practice. 
We summarise here the most important questions, and in Sec.~\ref{sec:quantum_network_medicine} we will discuss how quantum computers can help address some of them.

\subsubsection{Incompleteness of the interactome}
At a very general level, one of the main overall issues is the incompleteness of the molecular interactome databases.
Despite significant advances in systematic protein–protein interaction mapping technologies, it is estimated that the current coverage only includes about 20\% of all potential pairwise interactions \cite{Bonetta2010}.

Given that many network medicine techniques rely upon our knowledge of this network, as explained above, data incompleteness is a major issue. 
In fact, such a high level of incompleteness may potentially hinder the detection of small disease modules, since they are likely to be fragmented in the current molecular interactome model~\cite{menche_uncovering_2015}.
Computational predictive mapping approaches providing reliable de novo PPI detection could help to complement the data obtained through high-throughput experiments.
In addition, link prediction methods~\cite{kovacs_network-based_2019}, extensively employed in network theory, may be used to select computationally the most likely missing interactions and, therefore, guide experimental validation.

\subsubsection{Computational power}\label{sec:comppower}
Network medicine analysis requires handling large amounts of data to uncover its underlying structural properties, and it is, therefore, computationally demanding both in terms of memory and run-time. 
Regarding DDD, it is reasonable to assume---as data suggest---that drugs are generally not selective, so biologically active small molecules tend to have a high degree of promiscuity \cite{Chartier2017}.
This poses a challenge for medicinal chemists, as it demands characterising and fine-tuning complex chemical profiles of compounds (i.e., the list of their physico-chemical properties), to then identify those with the most promising biological profiles for the modulation of the disease network while minimising undesired toxicity and side effects \cite{Paci2022}.
For this to happen, one needs powerful tools for combinatorial and network search algorithms, as well as efficient methods for predicting the compounds' biological profiles and their effects on the disease network. 

\subsubsection{Beyond machine learning}\label{sec:BeyondML}
Machine learning (ML) has become a useful tool of modern biology and is currently part of several medical investigative workflows.
A number of ML applications have been successfully demonstrated in systems biology and pharmacology, including the prediction of protein binding interactions~\cite{wang_topology-based_2020}, the inference of gene regulatory networks~\cite{razaghi-moghadam_supervised_2020}, the identification of metabolic functions~\cite{shah_review_2021}, the discovery of key transcriptional regulators involved in several diseases~\cite{sonawane_understanding_2017}, and protein folding~\cite{jumper_highly_2021}.

Notwithstanding the many successes of ML and the new avenues that they have opened, there still exist serious difficulties, such as the need for an incredible amount of data to train the algorithms.
Moreover, one should remember that many advanced ML approaches work as ``black boxes'' from the outcomes of which it is difficult to infer specific feature correlations and provide biological interpretations.
For this reason, they are not generally suitable for the identification of mechanistic models.

\subsection{Network medicine for drug discovery and development}\label{sec:NetMedDDD}
Despite the significant increase in investment in biomedical and pharmaceutical research and development over the past 40 years, the annual number of FDA-approved treatments has not increased proportionally.
Many reasons have been proposed for this decline in productivity~\cite{sams-dodd_target-based_2005}.
A rather fundamental cause for such an impasse may be found in one of the core assumptions framing our approach to drug discovery.

Until recently, it was believed that the primary goal of drug discovery was the design of highly selective ligands acting on a single disease protein target.
The increase in the rate of drugs failing in late-stage clinical trials, however, challenges such an assumption.
The `one gene, one drug, one disease' paradigm has its origin at the intersection between genetic reductionism and the development of technologies that allowed for the isolation and characterisation of individual `disease-causing' genes.
This is based on Ehrlich’s philosophy of `magic bullets' targeting individual chemoreceptors governing specific diseases~\cite{kaufmann_paul_2008}.

Large-scale functional genomics studies have revealed that, under laboratory conditions, many single-gene knockouts do not affect strongly the phenotype~\cite{hopkins_network_2008}.
This robustness can be due to redundant functions and alternative compensatory signalling routes.
Remarkably, network medicine approaches have highlighted how such robustness of biological systems can be a consequence of the structure of the network.
As we have mentioned before, the scale-free nature of many biological networks results in systems that are resilient against random deletion of any one node but that are also critically dependent on a few highly connected hubs.

This property has profound implications for drug discovery;
instead of searching for the `disease-causing' genes, a much more promising strategy is to identify perturbations in the disease-causing module.
Network medicine analysis predicts that while the deletion of individual nodes often has no effect on the disease module, in order to perturb significantly the phenotype one may need to modulate multiple proteins.
As an example, dual knockouts in a number of model systems have shown that although the isolated deletion of two individual genes may have no effect, the simultaneous deletion of the two genes can be lethal or deleterious~\cite{ooi_global_2006}.

The \textit{three key challenges faced in the development of network medicine for DDD} are i) completing the interactome, ii) identifying a node or combination of nodes in a biological network, the perturbation of which results in a desired therapeutic outcome, and iii) discovering agents with the desired polypharmacology profile to perturb those nodes.
Quantum-enhanced network analysis approaches, combined with the existing methods, may lead to the disruptive leap needed to unlock a novel multiscale holistic DDD paradigm.

\section{Quantum Computing and Algorithms}\label{sec:QCA}
Quantum computers are devices that store and manipulate information at the level of microscopic constituents of light and matter (e.g., atoms, electrons, and photons).
Such microscopic particles obey the laws of quantum physics, in contrast to conventional computers where information storage and processing is described by the laws of classical physics~\cite{bennett_quantum_2000}.
While the fundamental unit of classical information is the bit which can take values 0 and 1, the fundamental unit of quantum information is called qubit (or quantum bit) and can be in a so-called quantum mechanical superposition of the states 0 and 1.

During the last two decades, the field of quantum information and computing has progressed to the point of demonstrating theoretically that quantum computers are able to harness quantum effects, such as quantum entanglement and superposition, to solve computational problems that are currently intractable with any conventional machine, including the most powerful supercomputers currently available~\cite{noauthor_40_2022}. 

\subsection{Quantum advantage}
A very large number of applications of quantum computers have been identified in the last decades, and they have the potential to lead to groundbreaking advances in finance, logistics, health, energy, security, and many other fields.
For all of these applications, there exist quantum algorithms that, in theory, prove advantage with respect to the best known classical algorithms~\cite{montanaro_quantum_2016}.
The holy grail of near-term quantum computing is to prove experimentally that existing small and imperfect quantum computers can already solve useful problems that conventional computers cannot~\cite{bharti_noisy_2022}.

We refer to \textit{useful quantum advantage} when a quantum device can beat a conventional computer in solving a problem that is useful in an academic sense, i.e., to advance science; we instead talk about \textit{practical quantum advantage} when it is ``practically useful'', that is, helpful for an industrially relevant use case.
Practical quantum advantage is much more difficult to achieve because of the general complexity of potentially relevant use cases, and it will probably only be realised after useful quantum advantage is achieved.
Hence, useful quantum advantage may be regarded as a prerequisite to practical quantum advantage. 

In order to program a quantum computer, it is necessary to develop new types of algorithms and software, given that they can only offer computational advantage if quantum phenomena are exploited.
Quantum computers without quantum algorithms are useless devices.
Most of the quantum algorithms that have been discovered assume that quantum computers are ideal machines, i.e., unaffected by noise-inducing errors.
They also work under the assumption that we can perform ideal read-out measurements to obtain the result of the computation.
Current implementations of quantum computers are very far from these conditions, which is why no demonstration of the usefulness of quantum computers for relevant use-cases exists to date. So far, quantum advantage has been demonstrated for a class of sampling problems with only academic relevance on superconducting quantum devices \cite{arute2019quantum} and photonic devices \cite{Zhong2020,Madsen2022}.

How far are we from useful quantum advantage, not just in terms of years, but most importantly in terms of research results?
We believe that achieving useful quantum advantage is not just a question of engineering or of technical improvements in the manufacturing of hardware components.
In addition to that, we need new algorithmic developments to manipulate and take advantage of the noisy data obtained with near-term quantum computers.
In other words, useful quantum advantage can be achieved with non-ideal hardware, but this demands the development of computational tools that compensate for the imperfections.
This goal is the main focus of Algorithmiq's current research and development efforts.

The actual difficulties in proving useful quantum advantage have come to light only during the last couple of years, as a result of the attempts to go beyond proof-of-principle experiments with existing quantum computers.
It is clear that there are at least two strategies and approaches to quantum advantage, which we discuss in the next section.

\subsection{Fault-tolerant vs near-term quantum computers}
The first strategy is to focus on the development of error-correction techniques, which would allow one to realise so-called \textit{fault-tolerant universal quantum computers}~\cite{paler_introduction_2015}.
These are devices where error correction techniques (akin to those routinely used in conventional computers) are incorporated, leading to the concept of a \textit{logical qubit}.
The implementation of a logical qubit involves many physical qubits and several operations that allow for the correction of random errors.
Fault-tolerant quantum computers would be able to run quantum algorithms in an ideal manner, hence automatically leading to useful and practical quantum advantage for all the aforementioned groundbreaking applications.
The problem is that there is still no scalable experimental implementation of logical qubits. The latest experiments show the implementation of one~\cite{egan_fault-tolerant_2021} or two logical qubits~\cite{erhard_entangling_2021,postler_demonstration_2022}, but large-scale demonstrations remain elusive.
For any practical applications, we would need at least hundreds or thousands of logical qubits.
This goal most likely will require a foundational research breakthrough.
It is expected that the timeline to achieve fault-tolerant quantum computing is 15 to 30 years.

The second strategy is to develop specific approaches for Noisy Intermediate Scale Quantum (NISQ) devices~\cite{preskill_quantum_2018}.
These include state-of-the-art quantum computers (with 5 to 100 qubits) and those that we expect to have in the next few years (with 100 to 1000 qubits).
The most straightforward application for useful (and practical) quantum advantage with NISQ devices is the simulation of other quantum systems, such as condensed matter systems (quantum materials) and molecules (quantum chemistry).
This is a ``natural'' use of quantum computers, since it is known that the exact simulation of many-body quantum systems with conventional computers is generally an exponentially hard problem~\cite{feynman_simulating_1982}.
It is believed that the reason why certain fields are experiencing major roadblocks is that we have reached the limits of what we can compute with today's supercomputers.
Examples of fields in which computational capabilities are a limiting factor are understanding high-temperature superconductivity, and simulating molecular systems in quantum chemistry.

\subsection{Hybrid algorithms for near-term quantum computers}
Developing algorithms for NISQ devices poses different challenges because one needs to face upfront the limitations imposed by noise and the relatively small number of qubits currently available in existing systems.
This is why all the algorithms for NISQ devices are hybrid, that is, they contain both a quantum part (algorithm run on a quantum computer) and a classical part (algorithm run on conventional computers)~\cite{mcclean_theory_2016,bharti_noisy_2022}.
The leitmotiv is that one wants to minimise the usage of the quantum computer to tackle only that part of the computational problem that really needs it, outsourcing all the classically computable tasks to the classical computer.
Mastering the proper balance between the two requires very specific competencies, the most important of which is understanding how the noise acts on quantum systems and how to reduce it or exploit it (a subfield of quantum physics known as open quantum systems).
The first example of useful quantum advantage will likely be obtained with NISQ devices.
It is also believed that NISQ devices will provide the first example of practical quantum advantage.
The timeline for achieving this goal is estimated to be 3 to 5 years. 

The main properties of quantum hardware that play a role in achieving quantum advantage are: 1) the number of qubits, 2) resilience to noise, both in terms of coherence times and of measurement and gate imperfections, 3) connectivity or topology of the device (which defines which pairs of qubits we can manipulate jointly by applying entangling gates), and 4) duration of the main gate operations.
There have been several attempts to introduce figures of merit for quantifying the performance of quantum hardware.
One of the first, introduced by IBM, is the quantum volume, which combines the first 3 factors into a single number~\cite{cross_validating_2019}.
The larger the quantum volume, the better the device is.
However, these single-number figures of merit, by construction, cannot capture all the complexities involved in the imperfections of the hardware.
In practice, one generally needs to assess all the factors from 1) to 4) and understand what is the dominant limitation for a given algorithm of interest on a case-by-case basis. 

One crucial point needs to be underlined: any quantum algorithm using less than about 50 qubits is simulable on a classical computer.
Therefore, the number of qubits is a necessary, but not sufficient, condition for demonstrating quantum advantage.

\section{Quantum Network Medicine}\label{sec:quantum_network_medicine}
In the previous sections, we have presented an overview of two emerging fields that promise a radical shift of perspective and enormous societal impact.
Remarkably, they seem to complement each other in providing an answer to either their current challenges (in network medicine) or their pressing needs (for quantum computing).
In this last section, we bridge the gap between the two by describing how quantum algorithms and quantum methods can be used within the framework of network medicine to either provide \textit{quantum enhancement} or \textit{quantum advantage}.
With the first term, we refer to any type of improvement in conventional network medicine approaches, including what is provided by the so-called \textit{quantum-inspired algorithms}, i.e., algorithms that are based on quantum phenomena but may still be run on conventional computers.
By quantum advantage, instead, we mean the most disruptive applications of quantum computing, namely those that cannot be achieved by any classical computer.
In the following two subsections, we focus on an example where quantum algorithms for near-term quantum computers can lead to quantum advantage, complementing network medicine in what ultimately is a quantum problem: the quantum mechanical simulation of the binding between a small molecule (ligand) and the biologically active site of a protein target, and its potential application to drug discovery.
Finally, in the last subsection, we give a brief overview of a number of quantum and quantum-inspired algorithms aimed at addressing disease-gene identification, the incompleteness of the interactome, protein folding and modelling, and search problems on structured databases. 

\subsection{Near-term quantum chemistry simulations}
One of the most promising applications of near-term quantum computers is simulating quantum systems.
Simulating a system can loosely be defined as using an alternative device (i.e. not the system itself) to predict the system’s behaviour or properties. 
Simulating quantum systems is classically very hard, as it requires dealing with mathematical objects, such as quantum states, of dimensions exponentially large in the system size.
This, in turn, demands exponential amounts of memory or compute time, and, hence, limits our ability to simulate large, or even moderate-size, systems in general.
While many techniques have been developed to avoid the need to deal with these intractable entities, they do not work in arbitrary situations, and, indeed, many relevant quantum systems, including molecular ones, remain beyond computational reach.

Quantum computers, which in fact were initially conceived for simulation, offer a much more direct and general way to overcome this limitation: since they are quantum systems themselves, they naturally exhibit the same phenomena that makes quantum systems difficult to simulate~\cite{feynman_simulating_1982,lloyd_seth_universal_1996}.
For example, while classical computers must store the description of the quantum system in a classical memory, a quantum simulator can be prepared in such state (or in a mathematically equivalent one), serving at one level as an analogue-equivalent of the system.
While the last sentence is a blunt oversimplification of the enormous advances in the field of algorithms for quantum simulation, it arguably summarises the spirit of most near-term approaches.

As explained in Sect.~\ref{sec:QCA}, near-term quantum computers cannot sustain very long calculations before the detrimental effects of noise become too significant.
Consequently, many of the elaborate algorithms conceived for fault-tolerant quantum computers cannot yet be applied.
However, even if not completely devoid of noise, a well-performing, large enough near-term quantum processor can be driven into a state that no classical computer can simulate, which can be a very valuable resource.
In particular, it is possible to map mathematically the quantum state of the electrons in a molecule and the quantum state of the qubits in the device to one another.
Classically inaccessible molecular states, such as the often sought ground state (the lowest energy state of the molecule), can then be encoded in the quantum machine, enabling us to extract the relevant quantities to predict its chemical behaviour.
This conceptually appealing idea was the key insight behind the celebrated Variational Quantum Eigensolver (VQE) algorithm, a seminal work that sparked this whole line of research~\cite{peruzzo_variational_2014}.

This approach, however, poses several significant challenges.
Given the relatively small number of qubits available, one cannot expect to be able to encode the full electronic state of the molecule.
Moreover, the presence of noise categorically prevents us from working with so-called pure states, such as the ground state, although low levels of noise can lead to almost pure states.
However, finding the right sequence of operations such that the resulting state of the processor approximates the ground state is far from trivial, especially considering that the number of operations itself must be minimised in order to reduce the effects of noise.
Finally, extracting the relevant information from the physical state of the device is a time-consuming process, as in order to access the information, quantum measurements yielding random outcomes need to be performed, with a precise evaluation demanding averaging over many repetitions.

The potential usefulness of near-term quantum chemistry simulations remains beyond reach as long as these issues are not addressed, and methods to tackle them have been proposed.
Computational chemists have developed many techniques to work with so-called active spaces, in which only the most complex part of the problem, the active space, is treated in a full quantum mechanical manner~\cite{hedegard_density_2015}.
These techniques can be adapted to the quantum realm by only solving for the active space quantum state on the quantum computer~\cite{rossmannek_quantum_2021}, hence allowing us to simulate very large systems that are classically inaccessible only due to a relatively small part of the problem.

In addition, further development of noise mitigation strategies, combined with the rapid improvement of hardware witnessed in the last few years, foreshadows high-quality results in the near future~\cite{endo_hybrid_2021,bharti_noisy_2022}.

Regarding the preparation of a good enough reference state on the quantum device, significant efforts have been devoted to understanding and overcoming the main hurdles, from unattainable circuit depths (that is, the count of non-parallelizable steps performed in a quantum computation) to the infamous barren plateaus (regions of the parameter space where the gradient is vanishing, halting the progress of the optimization procedure)~\cite{sack_avoiding_2022}, and it is becoming increasingly clear that iterative methods in which the circuit is constructed adaptively present considerable advantages~\cite{grimsley_adaptive_2019,grimsley_adapt-vqe_2022}.

Similarly, the problem associated with the large number of repetitions required has also been addressed in multiple ways.
In particular, the adaptive measurement scheme introduced by Algorithmiq in collaboration with IBM in Ref.~\cite{garcia-perez_learning_2021} not only reduces the measurement cost nearly quadratically, but it even provides us with so-called informationally complete measurement data, which enables a myriad novel methodologies to tackle all the other aforementioned issues. Informationally-complete measurements have been recently implemented experimentally on near-term quantum devices by the Algorithmiq team \cite{garciaperez2021experimentally}.

In short, near-term quantum simulation requires working with imperfect machines, so one should try to optimise every single step of the process in order to maximise the benefit of using NISQ devices.
In particular, for real-world applications, it seems sensible to consider a mostly classical computational chemistry pipeline in which the NISQ machine is an auxiliary device complementing the simulations whenever a problem beyond the capabilities of the classical algorithms is met.

\subsection{Quantum simulation in the DDD context}

Among the many properties of a given molecule that one may attempt to predict, its binding affinity to specific target proteins is of particular relevance in the context of DDD.
Essentially, the binding affinity quantifies how likely it is for the binding between the molecule and the protein to occur if the molecule successfully diffuses to the protein.
This, in turn, determines how effective the specific molecule could be at, for instance, inhibiting a specific protein in the cell by occupying one of its binding pockets.
Ultimately, the problem of calculating binding affinities requires simulating the state of the electrons in the binding pocket when the ligand is located in it, a problem that can be classically daunting, as previously explained.

As we mentioned in Sec.~\ref{sec:NetMedDDD}, one of the main goals of network medicine for DDD is identifying a protein (or a combination of proteins) in the interactome the perturbation of which results in a desired therapeutical outcome.
However, even if the nodal proteins are successfully identified, finding a molecule that can perturb the network accordingly is not an easy task, especially when accurate chemical simulations are not available.
Indeed, a number of computational approaches (including classical and more recently quantum machine learning, see Sec.~\ref{sec:BeyondML}) assume that ligands interact with proteins (associated with a disease) within the binding sites if they are structurally complementary.
The goal is to find optimally binding ligands by searching through all viable chemical compounds and molecules consistent with certain physio-chemical principles and conditions.
However, quantifying binding affinities in terms of structural complementarity can give very inaccurate results. Furthermore, without insight into the (network-based) functional consequences of binding, even more precise binding energy estimates would do little to improve drug efficiency predictions.

We, therefore, believe that a purely data-driven approach like network medicine, combined with the precise quantum estimation of binding affinities (and other relevant quantities), has an enormous potential to improve the DDD landscape.
This goal motivates our quantum network medicine approach.
We envision a framework in which network medicine would identify, based on holistic biological assessment, the nodes (proteins) for which we wish to maximise the probability of ligand binding and the nodes for which we wish to minimize binding to optimize the therapeutic-toxic ratio.
Quantum mechanical simulations could then be used to identify the most suitable ligands matching the desired (poly)pharmacological profile.

In fact, a quantum network medicine approach may bring other substantial benefits beyond the aforementioned enhanced selection procedure for known compounds.
It is widely believed that the space of drug-like molecules is many orders of magnitude larger than the set of known compounds, which suggests the existence of much more effective ligands~\cite{von_lilienfeld_exploring_2020,schneider_rethinking_2020}.
However, exploring such a vast space is extremely challenging.
Recently, a novel algorithm to search for candidate compounds that minimise the binding affinity to a specific pocket in combinatorially large sets has been proposed and even tested on real hardware~\cite{barkoutsos_quantum_2021}.
Extending this algorithm to multiple binding pockets may, thus, enable finding new drugs with a specific polypharmacological effect deemed desirable by network medicine analysis.

\subsection{Quantum-enhanced methods}
The latest quantum computing roadmaps reveal remarkable progress made during the last few years.
Not only larger and more robust near-term quantum computers have been deployed, but also noise mitigation and error correction strategies have been developed.
Interestingly, as byproducts of theoretical and experimental investigation in these areas, several improvements to algorithms running on classical computers, inspired by quantum properties, have been reported.
We discuss below some examples of quantum-inspired algorithms that present some advantages even when classically run, but that will reach their full potential when fault-tolerant quantum computers become a reality.

\subsubsection{Quantum walks for diffusion-based algorithms and spatial search}
Random walks are extensively used in systems biology and network medicine to describe diffusion processes.
From a mathematical perspective, a random walk is a random process describing a path that consists of a succession of random steps on some mathematical space.
In network medicine the walk occurs on the graph of interest, generally part of the interactome.
As an example, in Ref.~\cite{lee_identification_2018}, it is shown how a random walk with restart on the interactome outperforms previous methods in identifying effective drug-target interactions.
This is due to the fact that, in general, diffusive approaches exploit global network topology information and, therefore, can give better performance than local methods in some tasks.
Another example of the use of random walks is that introduced in Sec.~\ref{sec:DiseaseModule} for the identification of disease-gene candidates and, subsequently, disease modules.

\begin{figure*}[t]
    \includegraphics[width=0.85\textwidth]{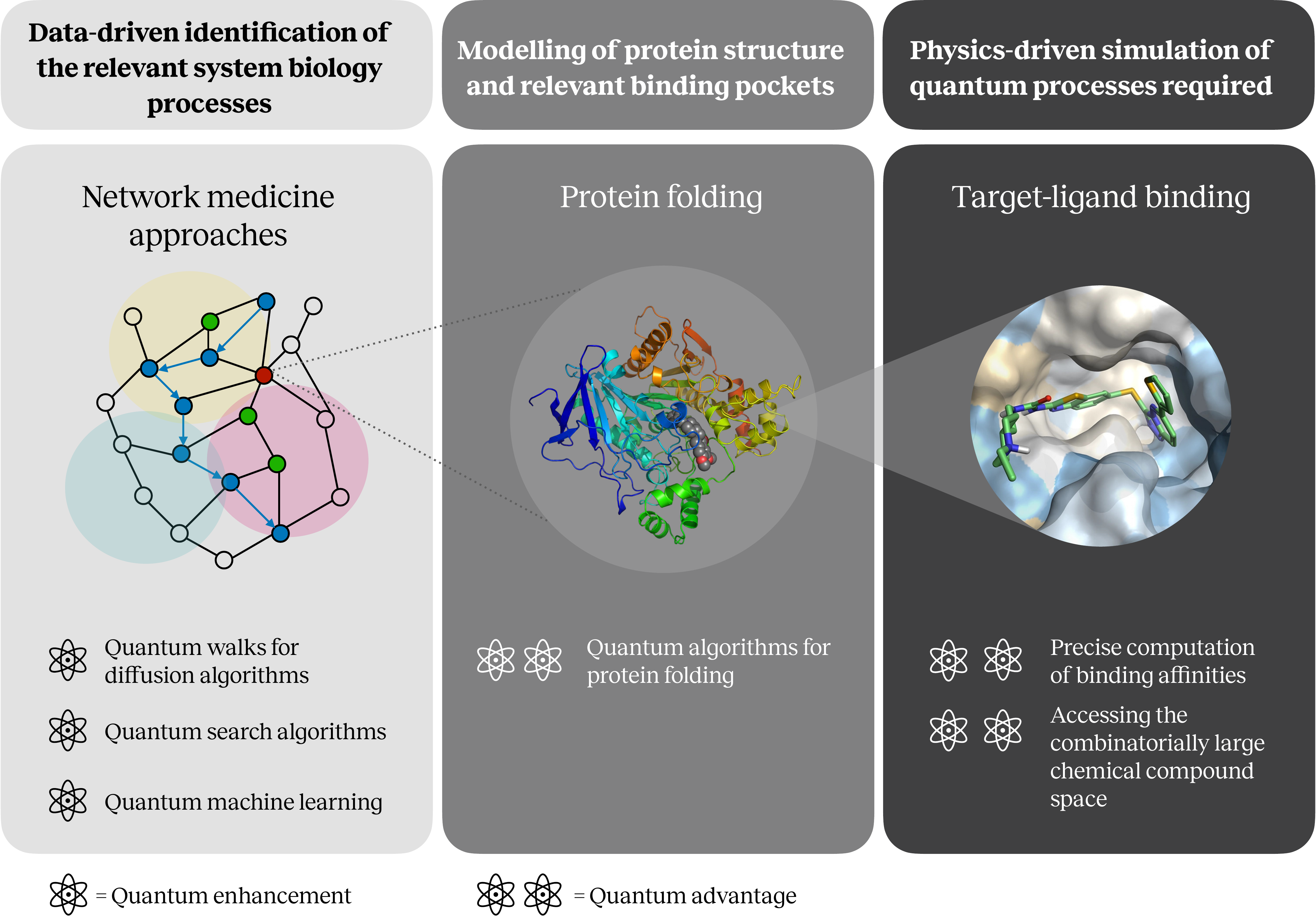}
    \caption{Multiscale holistic approach to drug discovery and development.
    Network medicine techniques, possibly enhanced by quantum algorithms, are used to determine the disease module and, consequently, to outline a holistically motivated biological profile to be met by candidate compounds.
    Structural models of the relevant proteins in the system are then built, via crystallographic reconstructions or with the help of (quantum) algorithms, to characterise their binding sites.
    Efficient quantum algorithms explore the vast chemical compound space to find molecular entities with the optimal binding affinity profiles.}
    \label{fig:fig3}
 \end{figure*}

A natural question then arises: can one improve these diffusive approaches using quantum methods?
The quantum version of a random walk is very well studied and is known as a quantum walk~\cite{kadian_quantum_2021}.
Classical random walks are purely Markovian processes with their variance increasing linearly with the number of steps.
By contrast§, quantum walks exploit phenomena that are inaccessible in the classical realm, such as quantum superposition and interference, contributing to faster (or slower) spreading.
The quantum speedup (or slowdown) stems from the constructive or destructive interference in the probability amplitudes of the different paths that the walker explores in the underlying dynamics.
The quadratic increase in the variance of the quantum walker (quadratic speedup) is a direct consequence of quantum coherence.
This quadratic speedup has been fruitfully exploited in the quest for improving various quantum algorithms~\cite{childs_exponential_2003,ambainis_quantum_2004, shenvi_quantum_2003, romanelli_quantum_2004} and machine learning~\cite{melnikov_predicting_2019}.
An interesting direction to explore is, therefore, whether the use of a quantum walk, instead of a classical random walk, would lead to a more efficient algorithm for the identification of disease-gene candidates or drug-target interactions, or would improve other diffusion-based applications on molecular networks.
Note that this does not necessarily imply that the networks themselves need to be quantum in nature.
This is in the spirit of quantum-inspired algorithms which, under some conditions and for certain problems, may perform better than their classical counterparts for classical problems.
As an example, one can show that continuous-time quantum walks (CTQW), i.e., the continuous limit of discrete-time quantum walks, can be used for network link prediction, yielding results competitive with state-of-the-art algorithms~\cite{goldsmith_link_2022}. 
This approach is, in turn, important in order to tackle the issue of the incompleteness of the interactome.

Finally, in Ref.~\cite{childs_spatial_2004}, it has been shown that CTQW for spatial search, i.e., the problem of finding a marked element in a structured database encoded by the topology of the network, is optimal for several graph topologies.
In other words, it provides quadratic speedup with respect to the corresponding classical algorithm.
Recently we have extended this investigation to the case of real complex networks~\cite{malmi_spatial_2022}.
We have used smaller network replicas obtained with a recent geometric renormalization method~\cite{garcia-perez_multiscale_2018}, and studied in this way the optimality of the algorithm as well as its scaling with respect to the size of the network.
Our results show that spatial search on real networks has, on average, considerably better scaling than the classical approach, but it does not reach the ideal quadratic speedup that can be achieved, e.g., in complete graphs.

These results indicate the potential utility of quantum tools in tackling current challenges arising from the inefficiency of combinatorial and classical network search algorithms (cf. Sec.~\ref{sec:comppower}).
For example, the quantum advantage shown for searches on tree networks using scattering random walks on quantum computers~\cite{koch_finding_2018} may enhance the identification of drug combinations used to treat complex diseases, proposed in Ref.~\cite{calzolari_search_2008}.

\subsubsection{Quantum algorithms for protein folding}
The protein folding problem, i.e., predicting the three-dimensional structure of a protein from its primary amino-acid sequence, has been intensively studied for over four decades.
This problem is known to be NP-hard, and is, therefore, considered unsolvable with classical algorithms.
Impressive results in predicting protein structures have been recently obtained by the neural-network-based  AlphaFold algorithm~\cite{jumper_highly_2021}, demonstrating accuracy competitive with experimental structures in a majority of cases and greatly outperforming other methods.
However, as with all machine learning approaches, AlphaFold does not reveal the mechanism or rules of protein folding, nor does it as reliably predict tertiary folds of random coil secondary structures.

A proof-of-principle experiment demonstrating a quantum algorithm for protein folding on near-term quantum computers has been recently put forward~\cite{robert_resource-efficient_2021}.
Specifically, a quantum variational algorithm is used to solve the folding of a polymer chain on a lattice, and it is shown that the algorithm has favourable scaling potentially enabling computationally efficient protein folding of large systems.
Experimental results on IBM quantum computers using 20 and 22 qubits demonstrate that, in principle, the algorithm functions as expected, and it, therefore, paves the way towards accessible and relevant scientific experiments on real quantum processors.

\subsubsection{Quantum machine learning for DDD}
Automated high-throughput screening technologies play a key role in drug development and discovery. Yet, the whole pipeline remains a long and expensive process.
To address this problem, computational molecular docking uses a number of machine learning approaches to identify target-binding compounds and to estimate the binding affinity between compounds and targets.
In Ref.~\cite{batra_quantum_2021,li_invited_2021}, the use of quantum machine learning algorithms for DDD has been investigated.
The quantum machine learning techniques presented include a hybrid quantum generative adversarial network to learn the patterns in molecular datasets and generate small drug-like molecules~\cite{li_quantum_2021}, a quantum classifier for protein pocket classification~\cite{li_invited_2021}, and a quantum variational autoencoder~\cite{khoshaman_quantum_2018} to generate a probabilistic cloud of molecules.
These methods are still affected by the poor performance of near-term quantum computers but may be useful in the future to enhance some of the steps of the DDD pipeline.
As mentioned in Sec.~\ref{sec:BeyondML}, however, such techniques are not helpful in understanding the biological functioning and deploying predictive models.
In this sense, they have limited scope with respect to the more ambitious goals and vision of quantum network medicine.

\begin{figure}[t]
    \includegraphics[width=\columnwidth]{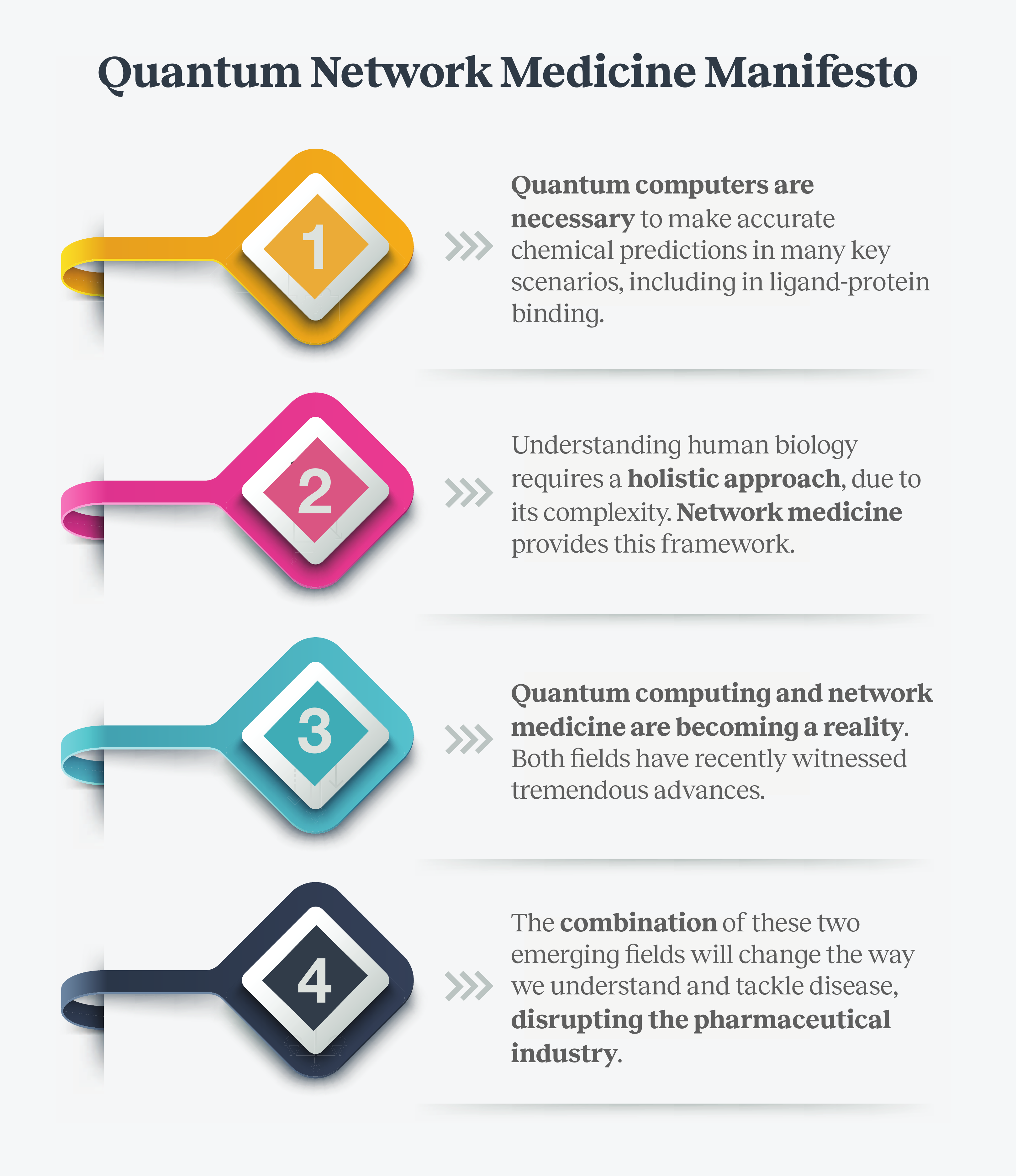}
 \end{figure}

\section{Final remarks}
We conclude this white paper by condensing our discussion in what we have denoted \textit{Algorithmiq's Quantum Network Medicine Manifesto}, consisting of four main points that highlight the importance and urgency of this new interdisciplinary endeavour. 

1.~\textit{Making accurate chemical predictions in many key scenarios, including in ligand-protein binding, is beyond reach for classical computers.}
Quantum computers present a natural solution to this problem and, in fact, quantum chemistry simulation is one of the most direct applications of quantum computing.

2.~\textit{Human biology cannot be effectively understood in terms of its parts in isolation.}
A complex systems, holistic approach such as network medicine is essential.
Indeed, network medicine has already proven its ability to provide valuable insights into disease causation, prevention, and treatment.

3.~\textit{Quantum computing and network medicine are becoming a reality.}
Both disciplines have witnessed astounding progress over the last decade, with a tempo that hints at an accelerated growth.
Quantum hardware has improved to the point of making possible the first experimental demonstrations of quantum advantage, albeit not useful or practical as yet, and it is now clear that useful and practical quantum advantage are only a matter of time.
In parallel, the field of quantum algorithms for quantum chemistry simulation has continued to provide solutions to the main hurdles posed by near-term quantum hardware imperfections.
This combination of hardware and algorithmic improvements promises to enable solutions to currently unsolvable chemistry problems in the near future.
In addition, owing to the rapid advance in both fronts, we can expect to be able to deal with increasingly large and complex chemical systems computationally, especially as we approach the fault-tolerant era.

In the case of network medicine, many successful applications have been achieved in the last four or five years, and we can expect many more significant results in the years to come.
In addition, network medicine will become increasingly powerful over time as a result of both the expansion of the available body of biologically relevant data, and of the increase in computational capabilities and methods.
In this regard, there is substantial room for improvement by using powerful network-theoretic tools, such as those developed in the field of network geometry~\cite{serrano_self-similarity_2008,krioukov_hyperbolic_2010,garcia-perez_mercator_2019,boguna_network_2021}, as well as quantum algorithms.

4.~\textit{The combination of these two emerging fields will change the way we understand and tackle disease.}
The high attrition rates in clinical trials that dominate current DDD pipelines can arguably be attributed, to a large extent, to an oversimplification of the underlying biology.
Network medicine is conceived to provide an integrated perspective of the relevant biological processes and, therefore, find more effective drug targets, as well as to foresee and minimise potential adverse effects.
However, even if network medicine analyses can unveil comprehensive biological profiles to be met by effective compounds, it is not possible to predict accurately the biological profile of compounds through purely conventional data-driven methods, especially for compounds for which no data are available---which is the case for the overwhelming majority of molecules in chemical compound space.
Therefore, such predictions need to be made from first principles, that is, by simulating quantum chemical processes.

Given these considerations, we envision a multiscale framework exploiting the unique strengths of quantum computing and network medicine, as well as their complementarity, as depicted in Fig.~\ref{fig:fig3}.
At the largest scale, network medicine techniques are used to identify the disease module and to produce a putative biological profile, which may also include information about drug-protein interactions that should be avoided.
Once the desired polypharmacological effect is understood, the structure of the relevant proteins must be determined, either computationally or via crystallography studies, in order to model their binding sites.
Finally, the chemical compound space is explored algorithmically, using efficient quantum algorithms tailored for finding molecular entities matching the biological profile.

If successfully implemented, both at the network medicine and quantum chemistry scales, an approach of this form would yield a transformative change in drug development and therapeutics.
On the one hand, it would open the gates to a combinatorially large space of candidate molecules.
On the other hand, the resulting molecules would present much higher success rates in clinical phases owing to the holistic, network-based assessment of their biochemical effects.
While this is a very ambitious program, it sets a concrete path to overcome the main challenges in drug development and effective therapeutics.

\section{Acknowledgements}
We acknowledge precious feedback on this manuscript from Joseph Loscalzo, and interesting and fruitful discussions with Joseph Loscalzo, M. Ángeles Serrano, Antonio Acín, and Vincenzo Cerullo during Algorithmiq's symposium ``At the Interface between Quantum Physics, Complex Networks, and Life Sciences'' held in Helsinki, 14 June 2022  \footnote{\url{https://algorithmiq.fi/network-medicine-symposium}}.

\bibliography{refs} 

\end{document}